\documentclass[aps,prl,reprint,superscriptaddress,nofootinbib,floatfix]{revtex4-2}

\usepackage{graphicx}
\enlargethispage{\baselineskip}
\usepackage{color}
\usepackage{latexsym}
\usepackage{amsfonts}
\usepackage[english]{babel}
\usepackage[colorlinks=True]{hyperref}
\usepackage[T1]{fontenc}
\usepackage[normalem]{ulem}
\usepackage{amsmath}
\usepackage{amssymb}
\usepackage{xcolor}
\usepackage{xspace}
\usepackage[utf8]{inputenc}
\usepackage{ucs}
\usepackage{txfonts}
\everymath{\displaystyle} 
\hypersetup{
    colorlinks=true,
    linkbordercolor={white},
    linkcolor={blue},
    citecolor={blue}
}
\graphicspath{{./}{figures/}}
\def\m87{{M87$^{\ast}$}\xspace}

\newcommand{\kharma}{\texttt{KHARMA}\xspace}

\begin{document}

\title{GRMHD accretion beyond the black hole paradigm: Light from within the shadow}
\author{Saurabh}
\email[Corresponding author: ]{saurabh@mpifr-bonn.mpg.de}\thanks{Member of the International Max Planck Research School (IMPRS) for Astronomy and Astrophysics at the Universities of Bonn and Cologne}
\affiliation{Max-Planck-Institut für Radioastronomie, Auf dem Hügel 69, D-53121 Bonn, Germany}

\author{Maciek Wielgus}
\email[Corresponding author: ]{maciek@wielgus.info}
\affiliation{Instituto de Astrofísica de Andalucía-CSIC, Glorieta de la Astronomía s/n, E-18008 Granada, Spain}

\author{Parth Bambhaniya}
\affiliation{Instituto de Astronomia, Geofísica e Ciências Atmosféricas, Universidade de São Paulo, IAG, Rua do Matão 1225, CEP: 05508-090 São Paulo - SP - Brazil}

\author{Elisabete M. de Gouveia Dal Pino}
\affiliation{Instituto de Astronomia, Geofísica e Ciências Atmosféricas, Universidade de São Paulo, IAG, Rua do Matão 1225, CEP: 05508-090 São Paulo - SP - Brazil}
\author{Andrei P.~Lobanov}
\affiliation{Max-Planck-Institut für Radioastronomie, Auf dem Hügel 69, D-53121 Bonn, Germany}
\author{Pankaj S. Joshi}
\affiliation{International Centre for Space and Cosmology, Ahmedabad University, Ahmedabad, GUJ 380009, India}

\date{\today}

\begin{abstract}
We present the first three-dimensional general relativistic magnetohydrodynamic simulation of sustained accretion onto a horizonless singularity in which matter reaches the central object rather than being accumulated outside of it or expelled in outflows. We consider a~Joshi–Malafarina–Narayan (JMN-1) spacetime, a~well-motivated black hole mimicker that arises from gravitational collapse with anisotropic pressure in general relativity, and adopt a compactness parameter for which the central singularity is null. We find that the system evolves into a sustained magnetically arrested disk state. For parameters appropriate to the low-luminosity active galactic nucleus M87*, we obtain an accretion rate of $\sim(3.0 \pm 0.5)\times 10^{-6} \dot{M}_{\rm Edd}$, in full agreement with estimates based on black hole models and, in particular, comparable to that of our reference Schwarzschild black hole simulation. Synthetic ray-traced images at $230\,{\rm GHz}$, computed using polarized general relativistic radiative transfer, are broadly consistent with the Event Horizon Telescope observations of M87*. We identify a key observational discriminant between a black hole and JMN-1: the presence of detectable brightness inside of the ``observable shadow" of JMN-1. This emission originates very close to the central singularity, in a region that would be hidden behind the event horizon in a black hole spacetime. Although this signature is beyond the reach of current observations, it falls within the projected imaging dynamic range of next-generation radio interferometric instruments, offering a robust test of the black hole paradigm.
\end{abstract}


\maketitle

\textit{\textbf{Introduction}}-- Black holes (BHs) are predicted by the vacuum equations of general relativity (GR). The fundamental solution, Schwarzschild metric, describes a~spherically symmetric static compact object possessing an \emph{event horizon}, a one way causal boundary cloaking the central singularity. This solution, as well as its rotating generalization Kerr metric \cite{Kerr1963}, carries theoretical pathologies which remain unresolved. In particular, no consensus resolution of the information paradox \cite{Hawking1976} exists despite decades of effort and the proposed solutions introduce further inconsistencies at the semi-classical level \citep{2013JHEP...02..062A}. The weak cosmic censorship conjecture \citep{singularity_theorem}, which posits that singularities are always cloaked by horizons, remains unproven and GR does admit collapse solutions terminating in horizonless configurations \cite{Christodoulou84, Ori1987}. Recent observational breakthroughs, including gravitational waves detection \citep{LIGO1}, stellar orbits monitoring at the Galactic Center \citep{GRAVITY1}, and horizon-scale imaging of M87* and Sagittarius~A* by the Event Horizon Telescope (EHT) \citep{EHT_M87_2019_I, EHT_SgrA_2022_I} have established broad observational consistency with the black hole solutions. Therefore, despite the fundamental theoretical tensions, the predictive success of GR in the strong-field regime consolidated the black hole paradigm as the standard framework for modeling compact astrophysical objects across the mass spectrum.

With the EHT and its continuous developments \citep{Johnson2023,Doeleman2023,Raymond2024} we are gaining an unprecedentedly detailed insight into the interactions between compact supermassive objects and their immediate environments of accreting plasma and ejected outflows. Thus, an observational window to test the black hole paradigm is opening and it motivates extensive studies of horizonless ultra-compact objects. BH alternatives explored in this context include wormholes \citep[][]{Wielgus2020,Vincent2021,Bambhaniya:2021ugr}, boson stars \citep[][]{Olivares2020,Vincent2021}, and horizonless singularities \citep[][]{Shaikh2019a,Kluzniak2024, Uniyal2025,Bambhaniya:2022xbz}, each representing a~qualitatively distinct topology resolving the horizon problem. By considering such alternatives, and then confronting them with the observational data, the black hole paradigm can be placed on a stronger empirical footing \citep{2019LRR....22....4C}.

In this letter, we study the JMN-1 spacetime \cite{Joshi2011}, a~notable example of a~horizonless compact object and a~strong black hole mimicker \cite{EHT_SgrA_2022_VI}. We consider JMN-1 because, unlike many alternatives, it arises as a physically motivated solution of gravitational collapse in GR, providing a well-defined setting to test whether the event horizon is observationally required. Using numerical general relativistic magnetohydrodynamic (GRMHD) simulations we compare accretion in JMN-1 and Schwarzschild spacetimes and find remarkably similar properties. We then propose a viable observational strategy to distinguish the two with future radio-interferometric observations.

\begin{figure}
    \includegraphics[width=\columnwidth]{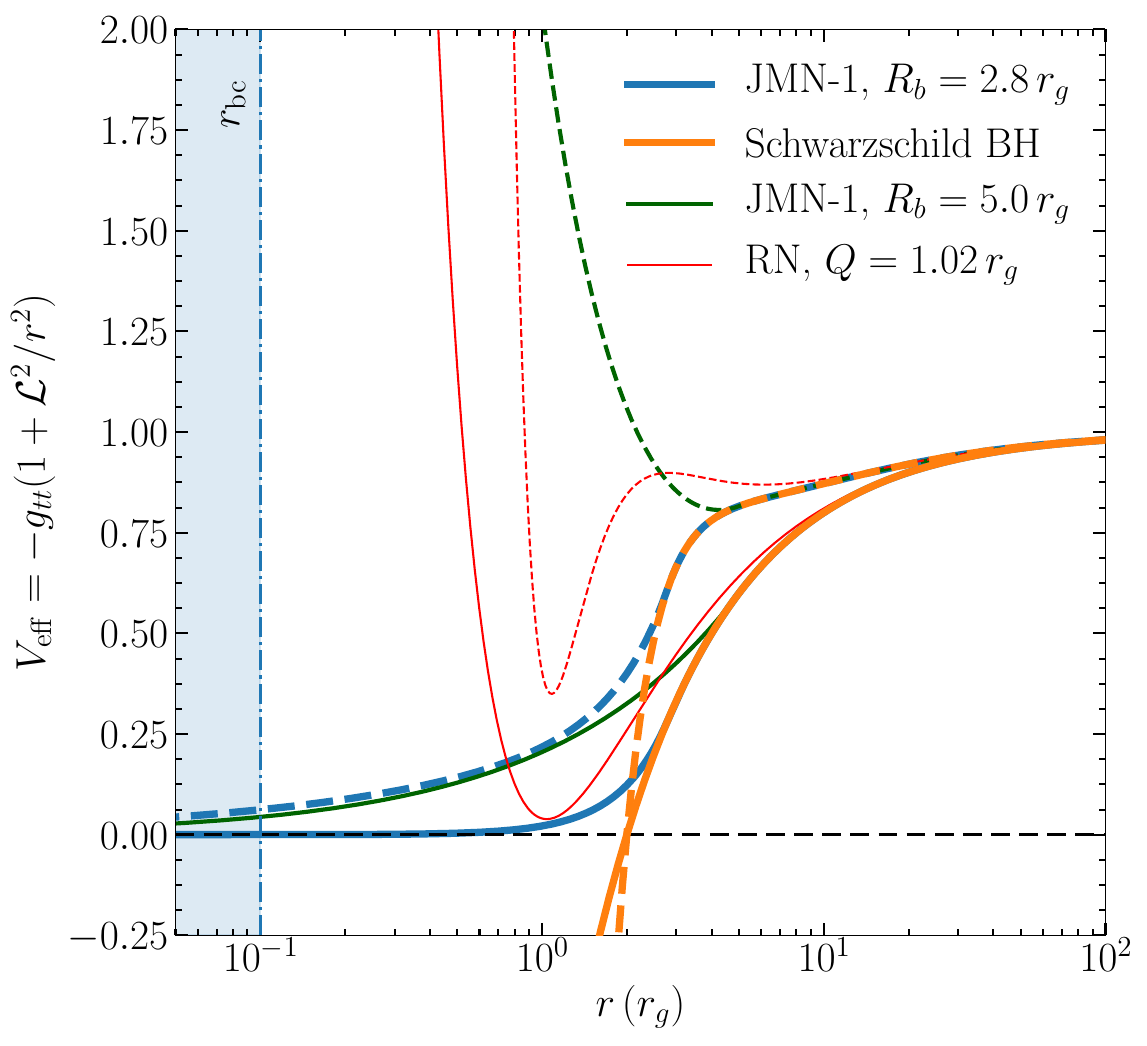}
    \caption{Comparison of the effective potential $V_{\rm eff}(r)$ for the metrics compared in this paper, Schwarzschild and JMN-1 with $R_b = 2.8\, r_g$, as well as for the Reissner-Nordstr\"{o}m (RN) naked singularity spacetime simulated in \cite{Kluzniak2024,Krajewski2025} and the static seed metric JMN-1 with $R_b = 5.0\, r_g$ of \cite{Uniyal2025}. We show the potential $V_{\rm eff}(r)$ for zero angular momentum massive particles $ \mathcal{L} = u_\phi = 0$ with continuous lines, and for $\mathcal{L} = 3\,r_g$ with dashed lines. RN naked singularity spacetimes (charge parameter $Q > r_g$) always indicate an infinite potential barrier preventing infall of particles. In JMN-1 spacetimes for $R_b > 3\,r_g$ the centrifugal potential barrier is present for all particles with $\mathcal{L} \neq 0$, rendering efficient accretion impossible. There is no potential barrier present in JMN-1 with $R_b < 3\,r_g$. The location of the inner boundary of our numerical simulations of JMN-1 $r_{\rm bc}$ is shown with a blue vertical line. }
    \label{fig:potentials}
\end{figure}

\begin{figure*}
    \includegraphics[width=0.99\linewidth]{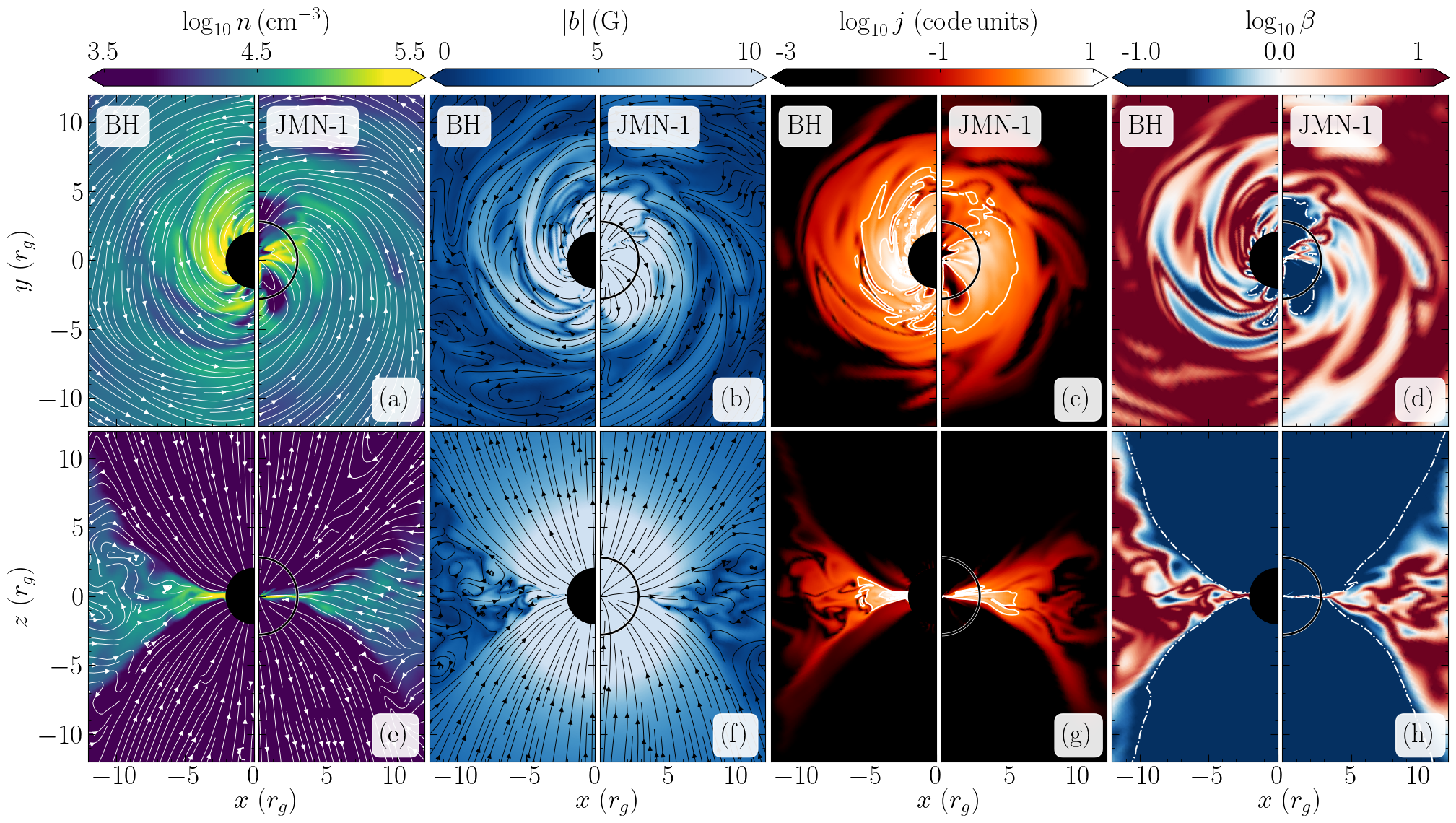}
    \caption{ Equatorial ($x-y$, top row) and meridional ($x-z$, bottom row) slices through a GRMHD simulation snapshot ($t=8000\,t_g$ for $t_g = r_g/c$). For each column the left panel shows Schwarzschild BH with a black filled semi-circle $r_{\rm h} = 2\,r_g$ indicating the event horizon. The right panels show JMN-1 with a semi-circle marking the matching radius $R_b=2.8\,r_g$. Maps of number density $n$ (panels a and e), magnetic field strength $|b|$ (panels b and f), emissivity proxy $j$ (panels c and g), and gas to magnetic pressure ratio $\beta$ (panels d and h) are shown. Solid white contours represent velocity flow lines in the number density $n$ panels (a) and (e), while black contours represent magnetic field lines in the magnetic field panels (b) and (f). The white contour in panels (c) and (g) indicates $50\%$ of the emissivity peak. The white dashed contour in panels (d) and (h) represents the magnetic to fluid energy density ratio $\sigma = 1$.}
    \label{fig:beta_momentum.png}
\end{figure*}

\textit{\textbf{Background spacetime}}---The horizonless JMN-1 spacetime arises as an equilibrium end state of inhomogeneous gravitational collapse in GR from regular initial conditions with nonzero tangential pressure and vanishing radial pressure \cite{Joshi2011}, unlike the idealized Oppenheimer–Snyder–Datt dust (pressureless and homogeneous) collapse that inevitably forms a BH \cite{Oppenheimer:1939ue,datt}. JMN-1 corresponds to a non-vacuum GR solution satisfying all standard energy conditions, forming a curvature singularity at $r=0$ 
and remaining free of closed timelike curves, thereby ensuring that causality is preserved \cite{Joshi2011,Joshi:2024gog}. JMN-1 is linearly stable against axial perturbations \cite{Pathrikar:2025ghp}, which, along with its dynamical stability, shows that the solution represents a genuine equilibrium configuration rather than a transient collapse stage.

Physically, JMN-1 appears as a spherically symmetric object of radius $R_b$, a compactness parameter of the solution, consisting of gravitating and otherwise non-interacting matter. A~requirement of subluminal speed of sound implies $R_b > 2.5\,r_g$, where $r_g = MG/c^2$ for a total (ADM sense) mass $M$ 
\cite{Joshi2011}. The metric inside the object is matched continuously to the vacuum Schwarzschild solution for $r \geq R_b$. The relevant JMN-1 metric tensor coefficients for $r \leq R_b$ are
\begin{eqnarray}
g_{tt} &=& -\left(1 - \frac{2\,r_g}{R_b}\right) \left(\frac{r}{R_b}\right)^{\frac{2\,r_g}{R_b-2r_g}} \ , \label{eq:gtt} \\
g_{rr} &=& \left(1 - \frac{2r_g}{R_b}\right)^{-1} \ .
\end{eqnarray}
For sufficiently compact configurations with $R_b < 3\,r_g$ JMN-1 spacetime admits Schwarzschild's photon sphere located at $r_\gamma = 3\,r_g$. Hence, such an object exhibits gravitational lensing structure similar to that of the Schwarzschild geometry \cite{Shaikh2019,Saurabh2024} and as a consequence the observable shadow of JMN-1 is consistent with the current EHT observational constraints \cite{EHT_SgrA_2022_VI,Vagnozzi2023}. Furthermore, even inspecting higher order images lensed around the photon sphere (the photon ring image structure), which is often proposed as a robust future test of the metric \cite{Johnson2020,Wielgus2021,Urso2025}, would necessarily yield results consistent with a Schwarzschild spacetime. To address this particularly challenging BH mimicker scenario, for our numerical considerations we fix $R_b = 2.8\,r_g$.

A somewhat less appreciated aspect of the studies of BH alternatives is whether they allow for efficient advection of energy. Both EHT horizon-scale targets correspond to low mass accretion rate, radiatively inefficient accretion flows, for which advective cooling plays a fundamental role for maintaining the energy balance \cite{Narayan1995}. In case of BHs, excess energy disappears behind the event horizon. For horizonless objects the energy may be removed by outflows, or may accumulate in the system. Either possibility qualitatively impacts the accreting system structure, its secular evolution, and the expected observational signatures. In many horizonless singularity spacetimes an infinite potential barrier prevents accretion. This is illustrated with several examples in Fig.~\ref{fig:potentials}, explaining the absence of inflow all the way to the central singularity in recent numerical studies \cite{Kluzniak2024, Krajewski2025,Uniyal2025}. A qualitative difference between the variants of JMN-1 spacetime in Fig.~\ref{fig:potentials} is related to the change of the $g_{tt}$ power-law index in Eq.~\ref{eq:gtt}, also associated with the transition of the singularity type from a~naked timelike for $R_b > 3\,r_g$ to a null type for $R_b < 3\,r_g$ \cite{BAMBHANIYA2023101215}. 

\textbf{\textit{Numerical simulations}}--- Numerical GRMHD constitutes the most physically complete framework for global simulations of compact accreting systems currently available at an acceptable computational cost. It allows us to model the interactions between matter and magnetic fields, fully accounting for special relativity as well as for the background curved spacetime, while preserving the fundamental laws of baryon number, energy, and momentum conservation. Magnetorotational instability  (MRI)~\cite{Balbus1991} develops natively in GRMHD simulations, enabling turbulent transport of angular momentum and facilitating accretion. An extensive library of numerical simulations and ray-traced images was assembled by the EHT collaboration to aid the theoretical interpretations of the M87* and Sagittarius~A* observations, with remarkable successes \cite{EHT_M87_2019_V,EHT_SgrA_2022_V, EHT_M87_2021_VIII,EHT_SgrA_2024_VIII}. For the vast majority of the existing global MHD simulations of relativistic accretion, Kerr spacetime background has been assumed, with several notable exceptions \cite{Mizuno2018, Olivares2020, Chatterjee2025, Dihingia2025, Uniyal2025}. In particular, simulations of accretion in horizonless singularity spacetimes constitute a very recent development \cite{Kluzniak2024,Cemeljic2025, Krajewski2025, Dihingia2025, Uniyal2025}, with only \cite{Dihingia2025, Uniyal2025} including magnetic fields and fully covariant formulation. Our simulation of JMN-1 with $R_b = 2.8 \,r_g$ is the first one demonstrating efficient accretion onto the central singularity in a horizonless spacetime, in contrast to the other simulations, in which matter is expelled from the innermost region. Thus, such a~background spacetime can be expected to mimic well the flow of energy around an astrophysical BH. The details of our numerical setup using \texttt{KHARMA} \cite{kharma}, modified to enable non-Kerr simulations, are given in the End Matter.

In Fig.~\ref{fig:beta_momentum.png} we compare snapshots from GRMHD simulations of Schwarzschild BH and JMN-1, started from identical initial conditions, demonstrating broad consistency for radii $r > R_b$ across the relevant quantities. At $r=R_b$, the metric is continuous and smooth in the sense of extrinsic curvature \cite{Shaikh2019}, but neither the metric connection nor the Ricci or Kretschmann curvature scalars are continuous there, as expected at an interface between vacuum and a self-gravitating material medium. Nonetheless, the numerical solution is well-behaving across the $r = R_b$ sphere, with weak shocks formation visible only through close inspection of quantities such as the gas pressure. These mild shocks are captured well by finite volume conservative numerical schemes such as those implemented in \texttt{KHARMA}. The flow is turbulent in both spacetimes and both solutions approach a magnetically arrested disk (MAD) state \cite{Narayan2003}. Strongly magnetized regions of low density, characteristic to MAD accretion, develop and appear more prominent in the JMN-1 simulation. This is particularly visible in a plot of gas to magnetic pressure ratio $\beta = p_{\rm gas}/p_{\rm mag}$ (panel d) and that of number density $n$ (panel a). Magnetic field geometry is very similar in both cases (panel f), typical to a MAD state \cite{SashaT2011}. In its innermost part, the accretion disk is geometrically thinner for JMN-1 than for a~BH, although both are thin in that region (panel e). We use the emission proxy $j$ (panels c and g; this quantity represents the scaling of the high-frequency limit of the thermal synchrotron emissivity, following Section 5.1 of \cite{Porth2019}) to demonstrate that most emission is expected to originate in the innermost region and close to the equatorial plane. Furthermore, in the case of JMN-1 significant amount of radiation originates very close to the singularity -- the analogous region is hidden by the event horizon in a BH spacetime. In the funnel region of JMN-1 outflowing velocity pattern develops, contrary to the BH case with inflow in the BH vicinity (panel e). However, both number density and radial velocity in the funnel region are very small in both simulations, so there is very little momentum transport differentiating the two solutions. 

We convert number density $n$ and the fluid frame magnetic field strength $\lvert b \rvert$ from the scale-invariant GRMHD code units to physical units by fixing M87* parameters \cite{EHT_M87_2019_I,EHT_M87_2019_VI,EHT_M87_2025_paperII}. As a~result, both simulations produce $n$ and $|b|$ (panels a, e and b, f of Fig.~\ref{fig:beta_momentum.png}) consistent with the constraints from \cite{EHT_M87_2021_VIII}. While the mass accretion rate is reduced for JMN-1, the suppression factor is only $\lesssim 2$. Importantly, both simulations result in mass accretion rates $ \dot{M} \sim(4-8) \times 10^{-4} M_{\odot}/$yr, or $\dot{M} \sim(3-5) \times 10^{-6} \dot{M}_{\rm Edd}$, where Eddington mass accretion rate $\dot{M}_{\rm Edd}$ is defined through Eddington luminosity $L_{\rm Edd}$ as $\dot{M}_{\rm Edd} = L_{\rm Edd}/0.1 c^2$. The reported values of $\dot{M}$ were averaged over $t \in (8-10)\times 10^3\,t_g$, see Fig.~\ref{fig:flux_evol} in the End Matter. These findings are very well consistent with the MAD Kerr BH models of M87* in the EHT library, which themselves vary between one another by a significantly larger factor \cite{EHT_M87_2019_V}.

\begin{figure}[ht!]
\includegraphics[width=\columnwidth]{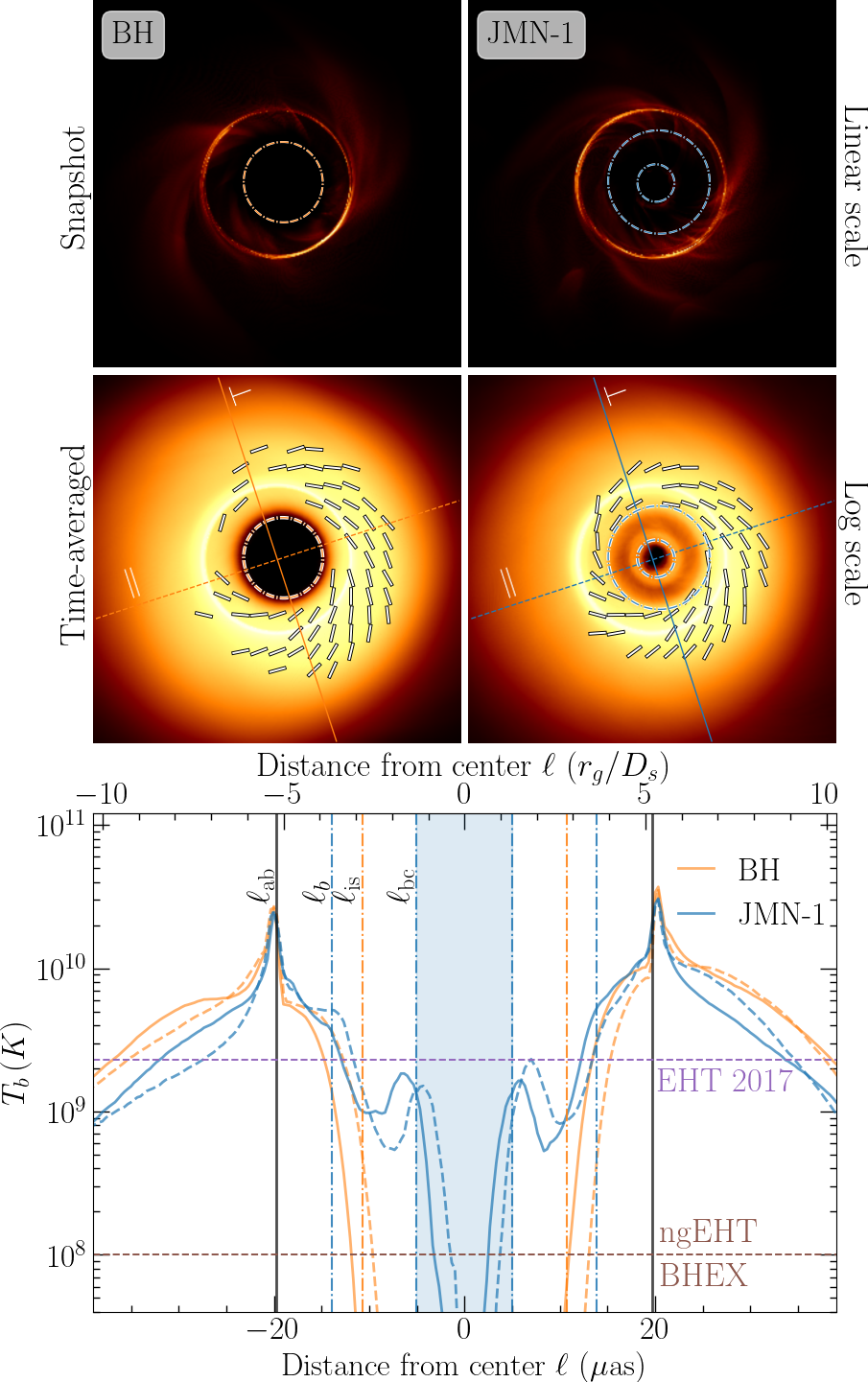}
    \caption{ \emph{Top row:} snapshot images $(t = 8000\,t_g)$  of a GRMHD accretion flow around a BH (left) and JMN-1 object (right), plotted in a linear colormap scale. The orange oval shape in the BH panel indicates the shape of the inner shadow $\ell_{\rm is}$, while the blue oval shapes in the JMN-1 panel correspond to $\ell_{\rm bc}$ the image of the inner boundary of the simulation in the equatorial plane, that is, an equatorial circle $r_{\rm bc} = 0.1 \,r_g$ (smaller one) and $\ell_{b}$ the image of the equatorial circle $r = R_b = 2.8 \,r_g$ (larger one). \emph{Middle row:}
    time-averaged images shown with a logarithmic brightness scale. Orientation of the polarization vectors is shown with ticks. Characteristic oval shapes and accretion-disk-spin-parallel ($\parallel$) and accretion-disk-spin-perpendicular ($\perp$) directions are indicated. 
    \emph{Bottom panel:} spin-parallel (dashed line) and spin-perpendicular (continuous line) cross sections through time-averaged images, comparing brightness temperature $T_b$ as a function of distance from the image center $\ell$. The vertical lines give the location of the apparent boundary $\ell_{\rm ab}$, inner shadow $\ell_{\rm is}$ of a BH, image of the JMN-1 matching radius $\ell_b$, and image of the inner boundary of the simulation domain $\ell_{\rm bc}$. These locations are exact for the spin-perpendicular slices (and for the face-on images).}
    \label{fig:avg_profile_m87}
\end{figure}


\textit{\textbf{Simulated images}}--- We used \texttt{ipole} \cite{Moscibrodzka2018} to perform a~full polarized relativistic radiative transfer calculations and to compute synthetic images of the two objects at 230 GHz, the observing frequency of the EHT. The images are strongly dominated by the synchrotron emission from relativistic electrons interacting with the magnetic field. We assumed characteristic parameters of M87*, namely, a distance $D_s = 16.9$\,Mpc, mass $M = 6.5 \times 10^9\, M_{\odot}$ and inclination angle $\theta = 163^\circ$ \cite{EHT_M87_2019_I,EHT_M87_2019_VI,EHT_M87_2025_paperII}. For the examples shown in Fig.~\ref{fig:avg_profile_m87} we additionally assumed the electron temperature parameter $R_{\rm high} = 50$ \cite{2016A&A...586A..38M,EHT_M87_2019_V}, adjusted the plasma density scale to match the EHT measurements of 230\,GHz flux density within 0.5-1.0\,Jy range \cite{Wielgus2020M87, EHT_M87_2019_III, EHT_M87_2025}, and matched the projected spin axis of the accretion disk with the projected jet axis, defined by the observed forward jet position angle of 72$^\circ$ clockwise from the north direction. In Fig.~\ref{fig:avg_profile_m87} we show snapshots and time-averaged images of the two objects. The colors represent brightness temperature, and the ticks in the second row panels additionally show the geometry of the linearly polarized emission. When viewed with the linear brightness scale, the two images are remarkably similar to one another, as well as to those from the M87* EHT imaging library \cite{EHT_M87_2019_V,EHT_M87_2021_VIII}. 

The prominent feature in the linear scale images is the location of the brightness peak -- the \emph{photon ring} \cite{Teo2003,Johnson2020,Wielgus2021}. This strong-lensing feature approximates the location of the \emph{apparent boundary}, which is the trace of the photon sphere $r_\gamma$ in the image plane. The apparent boundary is defined exclusively through spacetime geometry and it is located at the image-domain radius of $\ell_{\rm ab} = r_\gamma [-g_{tt}(r_\gamma)]^{-1/2} =  3 \sqrt{3}\, r_g/D_s$ for Schwarzschild spacetime, e.g., \cite{Bardeen1973}. The photon ring appearance is closely reproduced in the JMN-1 image, as the two spacetimes share identical structure above $r = R_b = 2.8\,r_g < r_\gamma$. Given the limited instrumental resolution, a robust observational detection of this sharp image feature currently remains elusive as it blends with the direct emission that spreads more flux over larger region of the image \cite{Johnson2020, LRR2025}. 

Inspecting the spin-parallel and spin-perpendicular image profiles shown in the bottom panel of Fig.~\ref{fig:avg_profile_m87} we see that the centroid of the brightness distribution does shift inward for the JMN-1 image. When integrated over the entire image ($\smallint  I \ell \,{\rm d} \Omega / \smallint  I \,{\rm d}\Omega$ for brightness $I$ and distance from the image center related to the photon's impact parameter $\ell = - k_{\phi}/k_t$), we find the difference of $\sim$10\%. This difference is not large enough to be readily distinguished from the impact of astrophysical setup, including uncertain plasma physics parameters such as $R_{\rm high}$ \cite{EHT_M87_2019_V,EHT_M87_2021_VIII} and from the mass and distance prior uncertainties \cite{EHT_M87_2019_VI}. 

First row of Fig.~\ref{fig:avg_profile_m87} presents the snapshot images in linear color scale and second row shows the time-averaged images of total intensity in the logarithmic color scale. In the second row images the presence of dim emission inside the apparent boundary $\ell_{\rm ab}$ is clearly visible. For a Schwarzschild BH image, under equatorial emission approximation, the brightness only extends inward to the \emph{inner shadow} \cite{Chael2021}, located at $\ell_{\rm is} \approx 2.85 \, r_g/D_s$ for a face-on observer. In that case the emission observed within the apparent boundary ($\ell_{\rm is} < \ell< \ell_{\rm ab}$) corresponds to photons emitted between the event horizon and the photon sphere ($r_{\rm h} < r < r_\gamma$). For JMN-1 the emission region extends all the way to singularity, as the event horizon is absent. Our study is limited by the inner boundary condition of the GRMHD simulation at $r_{\rm bc} = 0.1\,r_g$, the image of which is located at $\ell_{\rm bc} \approx 1.32\,r_g/D_s$ for the equatorial emission, face-on approximation, see Fig.~\ref{fig:avg_profile_m87} and the End Matter. The detailed structure of the image is sensitive to the accretion flow properties and the gravitational redshift of the spacetime. In the JMN-1 images from our GRMHD simulations we see a~mild rise of brightness near $\ell_{\rm bc}$, as gas density, temperature, and magnetic field strength all grow toward the singularity and so is the expected emissivity. While the appearance of this feature may be affected by the inner boundary condition treatment, our absorbing boundary condition should not artificially increase the emission in the simulation domain. In our spot-check tests we noticed that the feature depends also on the $R_{\rm high}$ parameter choice. Nonetheless, in all considered cases there is a dim but non-negligible brightness in JMN-1 images in the central image region, significantly more than in the BH images. The horizontal dashed lines in the right panel of Fig.~\ref{fig:avg_profile_m87} mark the approximate dynamic range of the EHT 2017 campaign, next-generation EHT (ngEHT) \citep{Johnson2023, Doeleman2023}, and Black Hole Explorer (BHEX) \citep{BHEX2024}. The projected dynamic range of ngEHT and BHEX is sufficient to probe the faint central emission predicted for JMN-1, which remains below the dynamic range of current EHT observations. In the End Matter we discuss lensing and redshift structure of JMN-1 spacetime. We also demonstrate the differences in the image structure between BH and JMN-1 for the analytic model of an equatorial accretion flow, free of the limitations related to the boundary condition treatment in GRMHD simulations. 

\textit{\textbf{Summary and discussion}}---We presented the first GRMHD simulation of accretion onto a JMN-1 null singularity horizonless object, demonstrating a sustained MAD state with an accretion rate comparable to that of a BH with the same mass. Hence, a singularity in such a spacetime acts as an efficient energy sink, similar to a BH's event horizon, allowing for an advection-dominated, radiatively inefficient accretion flow to form, as expected in objects such as M87* or Sagittarius~A*. This property is uncommon among horizonless alternatives to BHs, and renders such an object a particularly challenging BH mimicker.


Both Schwarzschild and JMN-1 spacetimes produce images consistent with current EHT observations of M87$^*$, yet differ structurally in the central image region. BH images indicate deep central brightness depression that can be understood approximately in terms of the inner shadow of the equatorial emission, related to the size of the event horizon. In the absence of the event horizon, intensity in the JMN-1 images extends inward below the inner shadow, and the effective inner boundary depends on detailed accretion flow properties. In our GRMHD example the image brightness at about half of the radius of the inner shadow reaches a few percent of the peak brightness. While confirming or excluding such a feature with the current EHT observations is impossible because of the low dynamic range of the image reconstructions, it should become possible with the planned EHT array upgrades within the next few years \cite{Johnson2023, Doeleman2023,LRR2025}, enabling a test of the BH paradigm. 

In this work we only discussed comparison between static spacetimes, Schwarzschild and JMN-1. Further studies of spinning models are necessary. Based on results of 
\cite{Chael2021} we expect Kerr characterization to be qualitatively similar to our findings, despite more expected foreground emission in the jet. For the JMN-1, finding geometry of the valid spinning generalization is not straightforward. An algorithmic solution was proposed by \cite{Kocherlakota2023}, but it is not guaranteed to obey GR equations with consistent equation of state, or at least with an admissible stress-energy tensor content. The geometry produced by \cite{Kocherlakota2023} was subsequently used for the GRMHD simulations of \cite{Uniyal2025}, but only in the naked timelike singularity regime $R_b > 3\,r_g$.

This pilot study motivates detailed further investigations of BH and JMN-1 observables, including polarimetric ones, quantifying differences between these distinct topologies for a larger space of model parameters. Our results highlight the importance of high image dynamic range, rather than extreme resolution, for the future tests of gravity. These developments will guide the implementation of the proposed BH paradigm test in the context of upcoming radio-interferometric instruments such as the next-generation EHT \citep{Johnson2023}.

{\it Acknowledgments}
The authors thank Cora Prather for helpful discussions related to \texttt{KHARMA} setup and Prashant Kocherlakota for helpful comments. This work was supported by the M2FINDERS project funded by the European Research Council (ERC) under the European Union's Horizon 2020 Research and Innovation Programme (Grant Agreement No. 101018682). MW is supported by a~Ramón y Cajal grant RYC2023-042988-I from the Spanish Ministry of Science and Innovation and acknowledges financial support from the Severo Ochoa grant CEX2021-001131-S funded by MCIN/AEI/ 10.13039/501100011033. EMdGDP acknowledges the support from the Brazilian Funding Agencies FAPESP (grant  2021/02120-0) and CNPq (grant 308643/2017-8). P. Bambhaniya, and EMdGDP acknowledge support from the São Paulo Research Foundation (FAPESP grant No. 2024/09383-4).


\appendix
\section{End Matter} 


\textit{Model setup}---We used \kharma\footnote{publicly available at \hyperlink{https://github.com/AFD-Illinois/kharma}{https://github.com/AFD-Illinois/kharma}} \citep[][]{kharma} (Kokkos-based High-Accuracy Relativistic Magnetohydrodynamics) to evolve ideal GRMHD equations. It is a performance-portable, \texttt{C++} implementation based on \texttt{iharm3D}, leveraging the Parthenon Adaptive Mesh Refinement Framework and the Kokkos programming model \cite[][]{CARTEREDWARDS20143202, 2021CSE....23e..10T} to run efficiently on CPUs and GPUs. It uses a second-order predictor–corrector scheme to step forward in time.

We implemented the JMN-1 metric in \kharma in Kerr-Schild like coordinates, following the approach described in \cite[][]{Kocherlakota2023} for non-vacuum spacetimes. Specifically, we used modified Kerr-Schild (MKS) coordinates $(x^0, x^1, x^2, x^3)$, with a resolution of $288 \times 128 \times 128$, consistent with the standard resolution adopted in the EHT GRMHD model libraries \citep{Porth2019, EHT_SgrA_2022_V}. The radial coordinate is exponential, $r = \exp(x^1)$, spanning from $r_{\rm bc}$ to $r_{\rm out} = 1000\,r_g$, increasing the density of grid 
zones at small radii. The polar coordinate is modified according to 
$\theta = \pi x^2 + \frac{1}{2}(1-h)\sin(2\pi x^2)$, with $h = 0.3$, concentrating grid zones toward the midplane ($\theta = \pi/2$) to better capture accretion disk physics, while widening zones near the poles. For JMN-1 the inner boundary is set at $r_{\rm bc}=0.1\,r_g$. Standard outflow boundary conditions are imposed at both the inner and outer radial boundaries. In the polar ($\theta$) direction, transmitting boundary conditions across the coordinate poles are applied. For JMN-1 the inner radial boundary excises the curvature singularity from the domain. Although no event horizon is present, outflow/absorbing conditions remain physically appropriate since the accretion disk inflow becomes strongly supersonic near the inner boundary, preventing causal communication with the exterior flow and thereby minimizing artificial reflections. The boundary condition is also well motivated by the spacetime potential structure, see Fig.~\ref{fig:potentials}. 

For both simulations we initialize a standard torus in hydrodynamic equilibrium \citep{Fishbone1976} with inner edge at $r_{\rm in} = 10\,r_g$ and pressure maximum at $r_{\rm max} = 20\,r_g$. We assume adiabatic index $\gamma = 5/3$. Small perturbations ($\pm~2\%$ randomization) are introduced in the fluid’s thermal energy to trigger MRI and incite accretion. The torus is threaded by a poloidal magnetic field loop, defined through a vector potential
\begin{equation}
    A_\phi = \max\left[ \left(\frac{\rho}{\rho_{\rm max}}\right)\left(\frac{r}{r_{\rm in}}\sin{\theta}\right)^3 \exp{\left( -\frac{r}{400}\right)} - 0.2, 0\right],
\end{equation}
with $A_r=A_\theta=0$ and plasma density maximum $\rho_{\rm max}$. The initial plasma beta, $\beta = p_{\rm gas}/p_{\rm mag}$, has a minimum value of 100. This setup enables sufficient magnetic flux to be advected toward the central object for the MAD state to develop. A complete description of the implementation of an equivalent torus can be found in Appendix A of \cite{patoka}.

\begin{figure}
    \centering
    \includegraphics[width=\columnwidth]{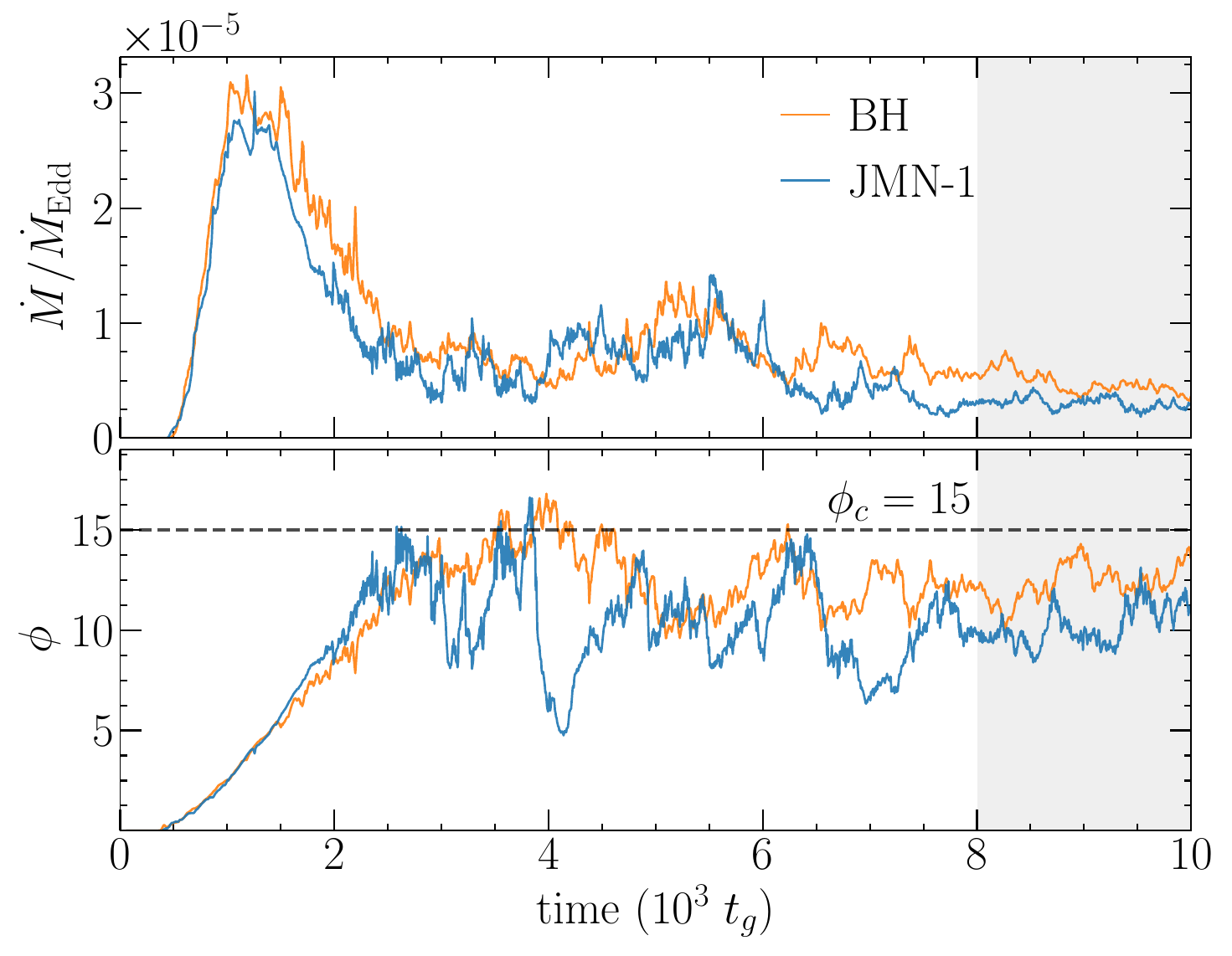}
    \caption{Evolution of mass accretion rate $\dot{M}$ normalized by the Eddington rate, and of normalized magnetic flux $\phi$ calculated at $r_{\rm h}=2\,r_g$ for BH and at $r_{\rm bc} =0.1\,r_g$ for JMN-1. For analyses we used the part of the simulation corresponding to a gray-shaded time range $t \in (8-10) \times 10^3 t_g$. The time-averaged $\dot{M}/\dot{M}_{\rm Edd}$ values are $\sim(4.6 \pm 0.9)\times 10^{-6}$ for BH and $(3.0 \pm0.5)\times 10^{-6}$ for JMN-1. The time-averaged values of $\phi$ are $12.4\pm1.0$ and $10.2\pm0.9$ for BH and JMN-1, respectively. }
    \label{fig:flux_evol}
\end{figure}

As a diagnostic, we compute mass accretion rate $\dot{M}$, and magnetic flux $\Phi_B$, through a sphere of radius $r$,
\begin{align}
    \dot{M} &= \int_{0}^{2\pi} \int_{0}^{\pi} \rho u^r \sqrt{\lvert\det(g)\rvert}\,{\rm d}\theta\,{\rm d}\phi, \\
    \Phi_{B} &= \frac{1}{2}\int_0^{2\pi} \int_{0}^{\pi} \vert B^r\rvert \sqrt{\lvert\det(g)\rvert}\,{\rm d}\theta\,{\rm d}\phi,
\end{align}
where $\rho$ is the fluid-frame gas density, $u^r$ is the radial component of the 4-velocity, $B^r$ the radial magnetic field component, and $\det(g)$ is the determinant of the metric $g_{\mu \nu}$. The level of magnetic saturation is characterized by the normalized magnetic flux $\phi=\Phi_B {\dot{M}}^{-0.5}$. The temporal evolution of $\dot{M}$ and $\phi$ are shown in Fig.~\ref{fig:flux_evol}. Both simulations reach a sustained MAD state characterized by $\phi_c \sim 15$. The accretion rate exhibits strong variability typical of MAD flows, while remaining comparable between the black hole and JMN-1 cases, and reaches a quasi-stationary state around $t = 6500\, t_g$. We also calculated the MRI quality factors $Q_{r,\theta,\phi}$, which specify the number of grid cells resolving the fastest growing MRI mode. We obtained mean values $\langle Q_{r, \theta, \phi}\rangle^{\rm{Sch.}}\gtrsim(30, 20, 6),\, \langle Q_{r, \theta, \phi}\rangle^{\rm{JMN-1}}\gtrsim(100, 20, 20)$, exceeding the minimum requirements \cite{Porth2019}.

\begin{figure*}
    \includegraphics[width=0.999\linewidth]{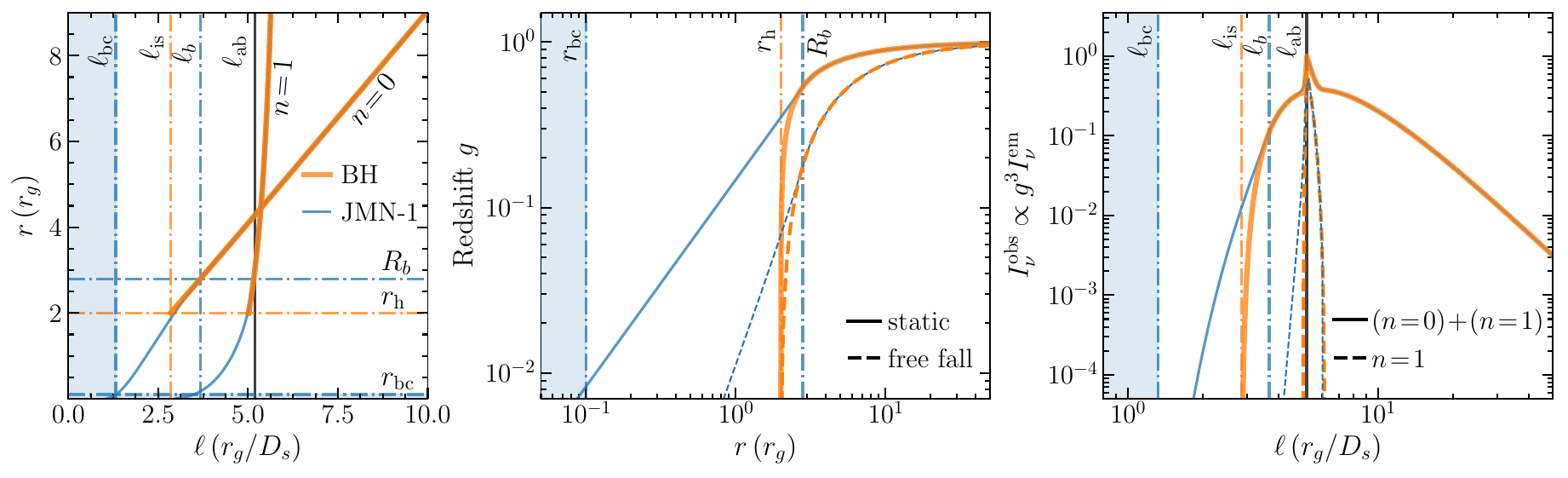}
    \caption{ \emph{Left:} transfer functions connecting photon's radial location in the image plane $\ell$ with the equatorial emission radius for the primary $n=0$ and secondary $n=1$ images for Schwarzschild BH (orange) and for JMN-1 with $R_b = 2.8\,r_g$ (blue). \emph{Middle:} gravitational redshift as a function of the emission radius in the two spacetimes for a static emitter (continuous lines) and a free-falling one (dashed lines). \emph{Right:} A~model of the observable brightness distribution $I^{\rm obs}_\nu (\ell)$ in the image, originating from equatorial free-fall emission, assuming profile of emission $I^{\rm em}_\nu \propto r^{-3}$. Profiles of secondary images are shown with  dashed lines.}
    \label{fig:lensing}
\end{figure*}
\textit{Image generation}--- We generated synthetic images by post-processing GRMHD simulation snapshots using \texttt{ipole}\footnote{publicly available at \hyperlink{https://github.com/AFD-Illinois/ipole}{https://github.com/AFD-Illinois/ipole}} \cite{Moscibrodzka2018,Prather2023}. The code numerically solves general relativistic polarized radiative transfer equations in a chosen background spacetime. The $230$\,GHz emission is synchrotron-dominated and we assume relativistic thermal spectral energy distribution of electrons \citep{EHT_M87_2019_V}. In the considered type of accretion flow electrons are thermally decoupled from ions \cite{Narayan1995}. As GRMHD evolves a single fluid temperature approximating the temperature of massive ions, we assign the electron temperature using a prescription \citep[][]{2016A&A...586A..38M,EHT_M87_2019_V} with an $R_{\rm high}$ parameter (fixed $R_{\rm low} = 1$). The field-of-view of images shown in Fig.~\ref{fig:avg_profile_m87} was set to $160~\mu{\rm as}$ (42\,$r_g/D_s$) with $(300\times300)~{\rm px^2}$ resolution.

\textit{Lensing and redshift in JMN-1 spacetime}---We consider a~mapping between the location in the image plane $\ell$  and the equatorial emission radius $r_{\rm em}$ under face-on viewing angle approximation, that is, a \emph{transfer function} \cite{Gralla2019}. In the first panel of Fig.~\ref{fig:lensing} we show transfer functions corresponding to the direct image $n\!=\!0$ and the secondary image $n\!=\!1$, the first component of the photon ring. We compare results for a Schwarzschild BH and for a JMN-1 horizonless spacetime with $R_b = 2.8\,r_g$. These are obtained by numerically integrating the photon deflection angle in each spacetime for assumed impact parameter $\ell$
\begin{equation}
\Delta \phi(r_{\rm em} ; \ell) = \int_{\infty}^{r_{\rm em}} \frac{k^\phi}{k^r} d \lambda \ ,
\end{equation}
where $k^\mu$ is the photon four-momentum, $k^\mu k_\mu = k^\theta = 0$ and $\ell = -k_\phi/k_t$ until reaching $\Delta \phi(r_{\rm em} ; \ell) = \pi/2$ for $n\!=\!0$ and $\Delta \phi(r_{\rm em} ; \ell) = 3\pi/2$ for $n\!=\!1$. The JMN-1 transfer function is identical with the BH one for the photons that remain in the Schwarzschild geometry, $r_{\rm em} > R_b$. However, the emission that reaches the distant observer is not limited to $r_{\rm em} > r_{\rm h}$, and instead corresponds to all $r_{\rm em} > 0$. The lensing very close to the singularity is extreme and a very small emission region $r_{\rm em} < 0.002\,r_g$ is mapped to a relatively large region of the image with $\ell < 1\,r_g/D_s$. We expect strong gravitational redshift to limit the detectability of the emission from the vicinity of the singularity. This is demonstrated in the second panel of Fig.~\ref{fig:lensing}, where the redshift is calculated for static and free-fall emitters. We define the redshift $g$ as the ratio between observed and emitted frequencies, calculated as
\begin{equation}
    g = \frac{\nu_{\rm obs}}{\nu_{\rm em}} = \frac{u^\beta_{\rm obs} k_\beta }{u^\alpha_{ \rm em} k_\alpha}
\end{equation}
for the four-velocity of the emitter $u^\alpha_{\rm em}$ and that of the observer at infinity $u^\beta_{\rm obs} = (1,0,0,0)$. For the static emitter $g =(- g_{tt})^{0.5}$ and we use the $n\!=\!0$ transfer functions for the $(\ell, r_{\rm em})$ mapping in the free-fall case. Full free-fall redshift formulas are given in \cite{Shaikh2019}. While the static redshift near the Schwarzschild horizon behaves like $g \propto (r-r_{\rm h})^{0.5}$, following Eq.~\ref{eq:gtt} we see $g \propto r^{1.25}$ dependence in the vicinity of the central singularity of our JMN-1. As a consequence, the impact of redshift is significantly stronger for the material approaching the center of JMN-1 object. Hence, the location of the matching radius $R_b$ is discernible in the log-scale image in Fig.~\ref{fig:avg_profile_m87} at an image domain distance $\ell_b$, with reduced brightness for $\ell < \ell_b$. Using transfer functions and redshifts, we construct a simple model of the observable image brightness profile corresponding to a free-fall equatorial emission, see the last panel of Fig.~\ref{fig:lensing}. Following the invariance of $I_\nu/\nu^3$ we compute the observable brightness distribution as $I^{\rm obs}_{\nu} \propto g^3 I^{\rm em}_{\nu}$. For illustration we assume the intrinsic emission profile $I^{\rm em}_{\nu} \propto r^{-3}$. While divergent at $r = 0$, the divergence of emission is mitigated by the strong JMN-1 redshift. We ignored effects of spectral index, disk thickness, and opacity, and we computed brightness distribution as a sum of primary and secondary images, neglecting higher order images $n \geq 2$. The image appears as a ring of approximately $\ell_{\rm ab}$ radius both in BH and JMN-1 case. However, there is a clear difference regarding the inner edge of the brightness distribution -- it necessarily terminates at the inner shadow $\ell_{\rm is}$ for a BH, but extends further inward for JMN-1, limited by an interplay of the intrinsic emission profile and redshift, thus demonstrating the same general result as our GRMHD simulations.

\bibliography{ref}{}

\clearpage 
 \newpage
\end{document}